\documentstyle[12pt,preprint,eqsecnum,aps,prd]{revtex}

\input epsf
\begin{document}

\preprint{\vbox{\hbox{MADPH-98-1086} 
                \hbox{FERMILAB-PUB-98/316-T}}}
 	  
\title{Adiabatic string shape for non-uniform rotation} 
\author{Theodore J. Allen}
\address{Department of Physics, Eaton Hall \\
Hobart and William Smith Colleges \\
Geneva, NY 14456 USA }

\author{M. G. Olsson}
\address{Department of Physics, University of Wisconsin, \\
1150 University Avenue, Madison, WI 53706 USA }

\author{Sini\v{s}a Veseli}
\address{Fermi National Accelerator Laboratory \\
P.O. Box 500, Batavia, IL 60510 USA}

\date{\today}

\maketitle

\thispagestyle{empty}
\vskip -20 pt

\begin{abstract}
It is well known that a straight Nambu-Goto string is an exact solution of
the equations of motion when its end moves in a circular orbit. In this
paper we investigate the shape of a confining relativistic string for a
general motion of its end. We determine analytically the shape of the
curved string to leading order in deviation from straightness, and show
that it reduces to an expected non-relativistic result. We also demonstrate
numerically that in realistic meson models this deviation is always
small. We further find that the angular momentum and energy are the same as
for the straight string, but that the curved string has a small radial
momentum not present in a straight string. Our results justify the common
assumption of straight strings usually made in hadron models.
\end{abstract}


\newpage

\section{Introduction}\label{sec:intro}

For some time flux tube models \cite{nambugoto,aft} and Wilson loop
expansion calculations \cite{brambilla} have been used to understand hadron
states.  The usual assumption is that the color field can be taken to lie
on straight lines connecting the quarks.  This ``straight line'' or
``rigid'' flux configuration is in the spirit of an adiabatic or
Born-Oppenheimer approximation that is assumed to hold at least for slowly
moving quarks.

Nesterenko \cite{nesterenko} points out that this adiabatic approximation
cannot be valid if the quark has angular acceleration.  The argument is
based on a classic theorem of ruled surfaces.  The Nambu-Goto string action
requires the string to sweep out a surface of minimal area.  Catalan's
theorem \cite{catalan} states that if the string is straight there is only
one minimal surface swept out; a helicoid.  This surface describes a meson
state consisting of uniformly rotating quarks.  As we will observe, a
straight string with the end moving radially also sweeps out a minimal
surface as long as the angular velocity is constant.  This solution does
not correspond to the motion of an actual meson since the quark plus string
angular momentum is not conserved.  However, one may take the point of view
that the string shape can be studied with arbitrary end motion, such as
caused by an arbitrary external force acting on the ends. The motion of the
string with quarks at the ends is then a special case.  In this paper we
use the rigidly rotating solution as our starting point and consider
angular acceleration of the endpoints as a perturbation.

A rigidly rotating string with its endpoint quarks moving in perfectly
circular orbits is not a realistic model for meson because quantum
mechanics requires there to be some radial motion, and by angular momentum
conservation, some angular acceleration as well.  The string must be curved
if it is to sweep out a minimal area while its endpoints undergo angular
acceleration.  We will discuss here the shape of the curved string and
conclude that small curvatures do not change its dynamics.  This result
holds even for relativistic string motion.

In section \ref{sec:ng} we define our notation for the Nambu-Goto action
and obtain the string wave equation, angular momentum and energy relations.
We also establish exact straight string solutions with arbitrary radial
motion but constant angular velocity.  An intuitive picture of a
non-relativistic string is developed in section \ref{sec:nrstring}.  The
relativistic shape equation and solutions for small angular acceleration
are established in section \ref{sec:relstring}.  We demonstrate numerically
in section \ref{sec:validity} that size of the string deformations is small
and therefore that the perturbative approach is sound.  The angular
momentum and energy for the curved string are considered in section
\ref{sec:constants} and our conclusions are given in section
\ref{sec:conclusion}.

\section{The Nambu-Goto-Polyakov string}\label{sec:ng}

The string action is proportional to the string tension $a$ and to the area
swept out by the string.  It is conventionally written in Polyakov
\cite{nambugoto} form as
\begin{equation}
S= -\,{a\over 2}\int d\tau\int_{\sigma_1}^{\sigma_2} d\sigma\, \sqrt{-h}\,
h^{ab} X^\mu\mathstrut_{,a} X^\nu\mathstrut_{,b}\,\, \eta_{\mu\nu}\ ,
\label{action}
\end{equation}
where $h_{ab}$ is a two-dimensional metric auxiliary field whose indices
run over $\tau$ and $\sigma$, and $h = \det(h_{ab})$.  $X^\mu(\tau,\sigma)$
is the string position and $X^\mu\mathstrut_{,a} \equiv \partial_a X^\mu$.  The
action is invariant under reparametrization of the coordinates $\tau
\rightarrow \tilde\tau(\tau,\sigma)$ and
$\sigma\rightarrow\tilde\sigma(\tau,\sigma)$, and under Weyl scaling of the
metric
\begin{equation}\label{weyl}
h_{ab}(\tau,\sigma)\rightarrow e^{\phi(\tau,\sigma)}\, h_{ab}(\tau,\sigma)\ .
\end{equation}
Because the metric $h_{ab}$ has no dynamics, we may solve its field
equations and substitute the result back into the action without changing
the equations of motion for the other fields.  The equations of
motion for $h_{ab}$,
\begin{equation}
X^\mu\mathstrut_{,a} X\mathstrut_{\mu,b} = {1\over 2} h_{ab}\, h^{cd} 
X^\mu\mathstrut_{,c} X\mathstrut_{\mu,d} \, ,
\end{equation}
and invariance under Weyl scaling (\ref{weyl}) allow us to set $h_{ab}$
equal to the induced metric
\begin{equation}\label{embmetric}
h_{ab} = X^\mu\mathstrut_{,a} X\mathstrut_{\mu,b} \ ,
\end{equation}
from which we see that (\ref{action}) reduces to, and is therefore
classically equivalent to, the standard square-root Nambu-Goto action.  

We will fix the coordinates $\tau$ and $\sigma$ by choosing $\tau$ to be
equal to coordinate time $t$, and $\sigma$ to run linearly along the length
of the string and have a fixed interval $(\sigma_1,\sigma_2)$ regardless of
the physical length of the string.  This simplifies the description of the
end-points of the string, but unfortunately precludes the use of
orthonormal coordinates.  A more detailed explanation of the
Nambu-Goto-Polyakov string, its geometrical aspects, and its relation to
QCD strings can be found in Ref.~\cite{barbashovnesterenko}.

The boundary conditions we take on $X^\mu(\tau,\sigma)$ are that one end is
fixed and the other is arbitrarily forced,
\begin{equation}
X^\mu(\tau,\sigma_1) = (t, {\bf 0})\ ,\quad X^\mu(\tau,\sigma_2) = (t, {\bf
x}(t))\ ,
\end{equation}
where ${\bf x}(t)$ is a prescribed function of time.  This simplifies our
presentation because we are only examining the piece of string from the
center of mass to one of the quarks.  It is a trivial generalization to
include quarks at both ends.  To include the dynamics of a (spinless) quark
on the end of the string, we would use its equation of motion,
\begin{equation}\label{quarkEoM}
\dot{p}^\mu = -\, a\, \sqrt{-h}\, h^{\sigma a} X^\mu\mathstrut_{,a} \ ,
\end{equation}
which is essentially Newton's second law with the tension in the string
providing the force.  The equation of motion (\ref{quarkEoM}) arises from
the variation of the boundary of the string when quarks are included from
the beginning.

Our purpose here is not to solve for the actual motion of the string--quark
system, which is a difficult problem, but rather to find the shape of a
string given a prescribed motion of its end.  Knowing the shape of the
string, and using Eq.~(\ref{quarkEoM}), we are in principle able to find
the motion of the string--quark system.  In practice, we use the
expressions for the total energy and angular momentum and quark momentum in
terms of the quark velocity to find the spectrum for the quantized
string--quark system numerically.  We demonstrate in
Appendix~\ref{app:cons} that Eq.~(\ref{quarkEoM}) is equivalent to the
conservation of energy, angular momentum and the quark mass-shell relation.

Planar motion is best described in terms of complex coordinates instead of
vector notation. In complex coordinates, the string position $X^\mu$ can be
written as
\begin{eqnarray}
X^\mu(\tau,\sigma) &=& (\tau, X^+(\tau,\sigma), X^-(\tau,\sigma),0)\ , 
\end{eqnarray}  
with
\begin{eqnarray}
X^\pm  &=& {1\over \sqrt{2}}(X^1\pm i X^2)\ .
\end{eqnarray}  
The metric $h_{ab}$, from Eq.~(\ref{embmetric}), is 
\begin{eqnarray}
h_{ab} &=& - X^0\mathstrut_{,a}X^0\mathstrut_{,b} + X^1\mathstrut_{,a}
X^1\mathstrut_{,b} + X^2\mathstrut_{,a} X^2\mathstrut_{,b} \nonumber \\
\label{metric} &=& - X^0\mathstrut_{,a}X^0\mathstrut_{,b} + 2\,\, {\rm
Re}\, (X^*_{,a} X_{,b})\ ,
\end{eqnarray}
where we adopt the notation $X\equiv X^+$.  

For simplicity, we consider here the fixed end at $\sigma_1=0$ and the
moving end at $\sigma_2=1$. Variation of the action with respect to the
position of the string yields the equations of motion
\begin{equation}
(\sqrt{-h}\,\,h^{ab}\, X^\mu\mathstrut_{,b})\mathstrut_{,a} = 0\ .\label{EoM}
\end{equation}
Once $X^\mu(\tau,\sigma)$ is known, the string four-momentum and angular
momentum are
\begin{eqnarray}
P_\mu &=& \int d\sigma\, \Pi_\mu(\tau,\sigma) =-\, a \int_0^1 d\sigma\,
\sqrt{-h}\,\, h^{\tau a}\, X_{\mu,a}\ , \label{Pmu}\\ 
J^3 &=& \int d\sigma\, X^{[1} \Pi^{2]} =-\, 2 a \int_0^1 d\sigma 
\,\sqrt{-h}\,\, h^{\tau a}\, {\rm Im}\,(X^* X_{,a})\ . \label{J}
\end{eqnarray}

The equations (\ref{EoM}) have an exact solution in which a straight string
rotates uniformly (constant angular velocity), but has an arbitrarily
changing length.  This is the solution that we will perturb to find the
string shape when its end undergoes angular acceleration.  Our ansatz is
\begin{eqnarray}
X^0(\tau,\sigma) &=& t \ =\ \tau\ , \nonumber \\
X(\tau,\sigma) &=& {\sigma R(t) \over \sqrt{2}}\, e^{i\omega t}\ . \label{ss}
\end{eqnarray}
From Eq.~(\ref{metric}), we find the metric tensor 
\begin{eqnarray}
h_{ab} &=& {\pmatrix{-\,\gamma^{-2}& \sigma R \dot{R} \cr
                     \sigma R \dot{R} & R^2\cr}}\ , \nonumber \\ \nonumber \\
\sqrt{-h} &=& R/\gamma_\perp\ , \\ \nonumber \\
\sqrt{-h}\, h^{ab} &=& \gamma_\perp\pmatrix{ -R           & \sigma\dot{R}\cr
                                            \sigma\dot{R} & R\gamma^{-2} \cr}\ , \nonumber
\end{eqnarray}
where we define 
\begin{eqnarray}
v_\perp &=& \omega R\ ,\nonumber \\
\gamma_\perp^{-2} &=& 1 - \sigma^2 v_\perp^2\ ,\label{gammas} \\
\gamma^{-2} &=& 1 - \sigma^2(v_\perp^2 + \dot{R}^2) \nonumber\ .
\end{eqnarray}
Substituting the above into the string equations (\ref{EoM}), we find that
they are exactly satisfied.  This result is the realization of Catalan's
theorem mentioned earlier, where the straight line to the quark sweeps out
a helicoid of fixed pitch but arbitrarily varying radius.  Again we
emphasize that this is a solution of the string equations (\ref{EoM}) only.
In a realistic meson, for which Eq.~(\ref{quarkEoM}) also holds for the
quarks on the ends, we must perturb Eq.~({\ref{ss}}) to find a solution
that takes into account the necessary angular acceleration of the ends.

For future
reference, we give here expressions for $J^3$ and $P_\mu$ of the straight
string.  From Eq.~(\ref{J}) we find that the angular momentum is given as
\begin{eqnarray}
J^3_{\rm straight} &=& -2 {a\over 2} \int_0^1 d\sigma \, \gamma_\perp
\,\left[ -R \,\, {\rm Im}(i\omega\sigma^2 R^2 + \sigma\dot{R}) + \sigma
\dot{R}\,\,{\rm Im}(\sigma R^2)\right] \nonumber \\ &=& a\omega R^3
\int_0^1 { \sigma^2 \, d\sigma \over\sqrt{1-\sigma^2v_\perp^2}} = {a R^2\over
2 v_\perp}\left[{\arcsin(v_\perp)\over v_\perp} - \sqrt{1 -
v_\perp^2}\right]\ . \label{Jstraight}
\end{eqnarray}
The corresponding energy and spatial momentum from Eq.~(\ref{Pmu}) are
\begin{eqnarray}
E_{\rm straight} &=& -a\int_0^1 d\sigma\,\, (-R\gamma_\perp) = a R\,
{\arcsin{v_\perp}\over v_\perp}\ , \\
P_{\rm straight} &=& {i\over \sqrt2} P_{\perp\, \rm straight}\, ,
\end{eqnarray}
where
\begin{eqnarray}
P_{\perp\, \rm straight} &=& a R^2 \omega \int_0^1 
{\sigma\,d\sigma\,\over\sqrt{1-\sigma^2 v_\perp^2}} \nonumber \\
&=& {a R \over v_\perp}\left[1 - \sqrt{1-v_\perp^2}\right]\ . \label{Pperp}
\end{eqnarray}
We observe that $P_{\rm straight}$ is purely transverse in direction.

It is also interesting to note the small velocity limits for these
quantities.  If we define the moment of inertia of a uniform string of
``mass'' $a R$ rotating about one end as
\begin{equation}
I = {1\over 3} (a R) R^2\ ,
\end{equation}
then the low velocity limits of $J^3$, $E$, and $P_\perp$ are
\begin{eqnarray}
J^3 &\rightarrow& I\omega\ , \nonumber \\
E &\rightarrow& a R + {1\over 2} I \omega^2\ , \\
P_\perp &\rightarrow& {1\over2}(a R)v_\perp\ .\nonumber
\end{eqnarray}
As one might expect, the string can be thought of as a rod of mass $a R$.

\section{Non-relativistic string shape }\label{sec:nrstring}

\begin{figure}[p]
\vspace*{2.0in}
\epsfxsize=\hsize
\hbox{\hskip 0 in \epsfbox{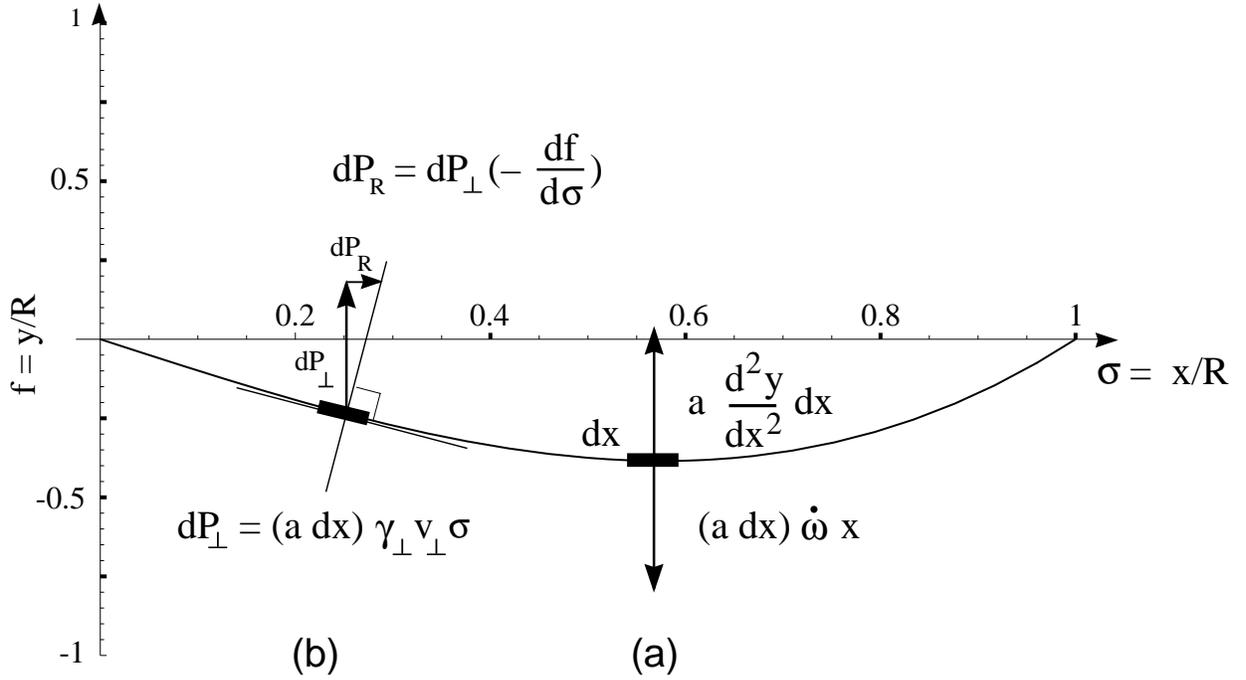}}
\caption{The solution to the non-relativistic string shape.  (a) Force
balance on a small element $dx$.  (b) Transverse and radial momentum of an
element.}
\label{fig:one}
\end{figure}

In the above we have demonstrated that a uniformly rotating string can
remain straight even if the length changes.  When the end has angular
acceleration the string must curve.  To gain intuition we consider a
quasi-Newtonian string of ``mass'' density $a$ which rotates with
instantaneous angular velocity $\omega$ about one end.  In the rotating
frame an element at distance $x=\sigma R$ is assumed to be in equilibrium
under two forces, the tension force
\begin{equation}
F_{\rm tension} = a\, {d^2 y\over dx^2}\, dx\  ,
\end{equation}
and the angular acceleration fictitious force
\begin{equation}
F_{\rm fictitious} = -\,(a\, dx)\, \dot{\omega} x\ ,
\end{equation}
as shown in Fig.~\ref{fig:one}.  The only other possible transverse force
is the Coriolis force due to the radial motion of the right end. However,
the element does not experience a Coriolis force since motion of the end
only creates more string and the notion of longitudinal velocity has no
meaning.  This type of motion can be thought of as ``adiabatic'' since the
resulting shape depends only on the end acceleration.  The force
equilibrium condition yields
\begin{equation}
{d^2y\over dx^2} = \dot{\omega}\, x\ .
\end{equation}
In terms of dimensionless variables, $f\equiv y/R$ and $\sigma = x/R$, the
above equation is
\begin{equation}
{d^2f\over d\sigma^2} = \dot{\omega}\,R^2\,\sigma\ .\label{nrstringshape}
\end{equation}
Using the end condition $f(\sigma = 0) = f(\sigma = 1) = 0$, we find the
non-relativistic string shape
\begin{equation}
f(\sigma) = -\,{\dot{\omega} R^2\over 6} \sigma (1-\sigma^2)\ , \label{cubic}
\end{equation}
which is illustrated in Fig.~\ref{fig:one}.

\section{Relativistic string shape}\label{sec:relstring}

In order to find the relativistic string shape, in this section we consider
the nature of small string deflections.  We again consider adiabatic
solutions generalizing the non-relativistic concept of the previous section
to the Nambu-Goto case.  We straightforwardly perturb about the straight
string solution (\ref{ss}) to take
\begin{eqnarray}
X^0(\tau,\sigma) &=& t\ , \nonumber \\
X(\tau,\sigma) &=& {1\over \sqrt2}\Big(\sigma R(t) + F(\sigma)\Big)\,
\exp{\Big[i(\omega t + \phi(t))\Big]}\ , \label{perturbX}
\end{eqnarray}
where the function $F(\sigma)$ is assumed complex.  We also assume that
$F(\sigma)$, $\dot\phi(t)$, $\ddot{\phi}\equiv\dot\omega$, and $\dot{R}$ are
small, and therefore we drop terms like $F\dot\omega$, $F\dot{R}$,
$\dot{\phi}\dot{R}$, etc.  Using Eq.~(\ref{metric}), we find in this
approximation
\begin{eqnarray}
h_{tt} &=& -\,{1\over \gamma^2} + 2\sigma\omega^2 R\,\, {\rm Re} F -
2\sigma^2\omega^2 R^2 \dot\phi\ , \nonumber \\
h_{t\sigma} &=& h_{\sigma t} = \sigma R\dot R - \omega R \,{\rm Im} F +
\sigma \omega\, {\rm Im} F^\prime\ , \\
h_{\sigma\sigma} &=& R^2 + 2 R\,\, {\rm Re} F^\prime\ \nonumber . 
\end{eqnarray}

The assumed string position (\ref{perturbX}) and the metric $h_{ab}$ above
must satisfy the equations of motion (\ref{EoM}) for the time and spatial
components 
\begin{eqnarray}
(\sqrt{-h}\, h^{a t})\mathstrut_{,a} &=& 0\ , \\
(\sqrt{-h}\, h^{ab}\, X_{,b})\mathstrut_{,a} &=& 0\ . 
\end{eqnarray}
Upon substitution, and with considerable but straightforward algebra, we find
that each equation is satisfied when $F(\sigma)$ satisfies
\begin{equation}
(1-\sigma^2 v_\perp^2) {d^2 ({\rm Im} F) \over d\sigma^2} + \sigma
v_\perp^2 {d ({\rm Im} F) \over d\sigma} - v_\perp^2 {\rm Im} F = \sigma
R^3\dot\omega\ .
\end{equation}
There are no constraints on the real part of $F$, which is a consequence of
the reparametrization invariance of the Nambu-Goto-Polyakov action
(\ref{action}).  In terms of a dimensionless quantity
\begin{equation}
f\equiv{\rm Im} F / R\ ,
\end{equation}
we find that the displacement from the straight string satisfies
\begin{equation}
(1-\sigma^2 v_\perp^2) {d^2 f \over d\sigma^2} + \sigma v_\perp^2 {d f
\over d\sigma} - v_\perp^2 f = \sigma R^2\dot\omega\ . \label{stringshape}
\end{equation}

For small rotational velocities $v_\perp$, the string shape equation
(\ref{stringshape}) reduces to
\begin{equation}
{d^2 f\over d\sigma^2} = \sigma R^2 \dot\omega\ ,
\end{equation}
which is identical to the non-relativistic result,
Eq.~(\ref{nrstringshape}).  The Nambu-Goto shape must then reduce to the
previous result, Eq.~(\ref{cubic}).

With the choice of independent variable
\begin{equation}
\xi = \sigma v_\perp\ ,
\end{equation}
the shape equation (\ref{stringshape}) reduces to the simpler form
\begin{equation}
(1-\xi^2) {d^2 f\over d\xi^2} + \xi {df\over d\xi} - f = \xi \left({\dot\omega
R^2\over v_\perp^3}\right)\ , \label{shapediffeq}
\end{equation}
whose exact analytic solution,
\begin{equation}
f(\sigma) = {\dot\omega R^2\over v_\perp^3} \left[ {1\over 2}\,\xi\,
\arcsin\xi + \sqrt{1-\xi^2} \right]\arcsin\xi + C_1 \xi +
C_2\left(\sqrt{1-\xi^2} + \xi\, \arcsin\xi\right)\ ,
\end{equation}
is discussed in Appendix~\ref{app:diffeq}.  The constants $C_1$ and $C_2$
are fixed by the end conditions that $f(\sigma)$ vanish at $\sigma = 0$ and
$\sigma = 1$.  The final result is
\begin{eqnarray}
f(\sigma) &=& {\dot\omega R^2 \over 6}\, {\rm shape}(\sigma)\ , \label{fpert}
\\ {\rm shape}(\sigma) &=& -\, {6\over v_\perp^3}\Bigg[ {1\over2}\sigma
v_\perp \left( (\arcsin v_\perp)^2 - (\arcsin \sigma v_\perp )^2 \right)
\nonumber \\ && +\, \sigma\,\sqrt{1-v_\perp^2}\arcsin v_\perp -
\sqrt{1-\sigma^2 v_\perp^2} \arcsin \sigma v_\perp\Bigg]\ . \label{shapefcn}
\end{eqnarray}

\begin{figure}[p]
\vspace*{2.0in}
\epsfxsize=\hsize
\hbox{\hskip 0.0 in \epsfbox{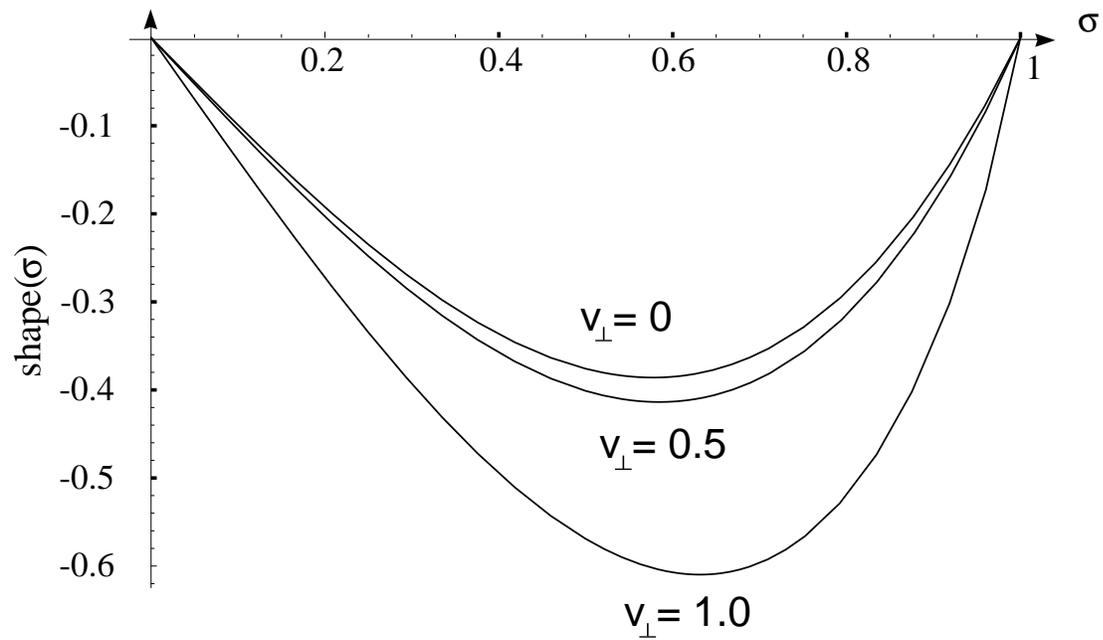}}
\caption{The relativistic string shape function given in
Eq.~(\protect\ref{shapefcn}) for various transverse velocities $v_\perp=\omega
R$.}\label{fig:two}
\end{figure}

By comparison of the above expression with (\ref{cubic}) we see that the
only difference from the non-relativistic result is in the shape function.
From the power series expansion it is straightforward to verify that
\begin{equation}
{\rm shape}(\sigma) \buildrel {{\tiny v_\perp \ll 1}} \over
\longrightarrow\, -\, \sigma\, (1-\sigma^2)\ ,
\end{equation}
and, hence, $f(\sigma)$ reduces to the non-relativistic limit.  In
Fig.~\ref{fig:two} we show $shape(\sigma)$ for non-relativistic,
intermediate, and fully relativistic speeds.  One can observe that even for
a very rapid rotation, $v_\perp = \omega R\rightarrow 1$, the string shape
does not change dramatically with respect to the non-relativistic result.

\section{Validity of the perturbative approach}\label{sec:validity}

As we have seen from Eq.~(\ref{fpert}), the actual size of the displacement
from the straight string is controlled by a factor of
${1\over6}\dot{\omega} R^2$.  Its magnitude for meson states can be
estimated using the heavy-light version of the relativistic (straight) flux
tube (RFT) model \cite{aft}.  Since the numerical solution of the model
provides us with a matrix representation for the $v_{\perp}$ operator in a
particular basis, it is convenient to rewrite $\dot{\omega} R^2$ as
\begin{equation}
\dot{\omega}R^2 = \dot{v}_{\perp}R-v_{\perp}\dot{R}\ .
\end{equation}
The above classical expression can be quantized by the appropriate
symmetrization procedure, and by promoting $v_{\perp}$ and $R$ to
quantum-mechanical operators, for which one can use $\dot{\Omega} =
-i[\Omega, \cal{H}]$. In this way, once the model is solved and matrix
representations for $v_{\perp}$ and the Hamiltonian $\cal{H}$ are found, it
is straightforward to compute the expectation value of $\dot{\omega} R^2$
for a given quantum state.

\begin{figure}[p]
\vspace*{2.0in}
\epsfxsize=\hsize
\hbox{\hskip 0.0 in \epsfbox{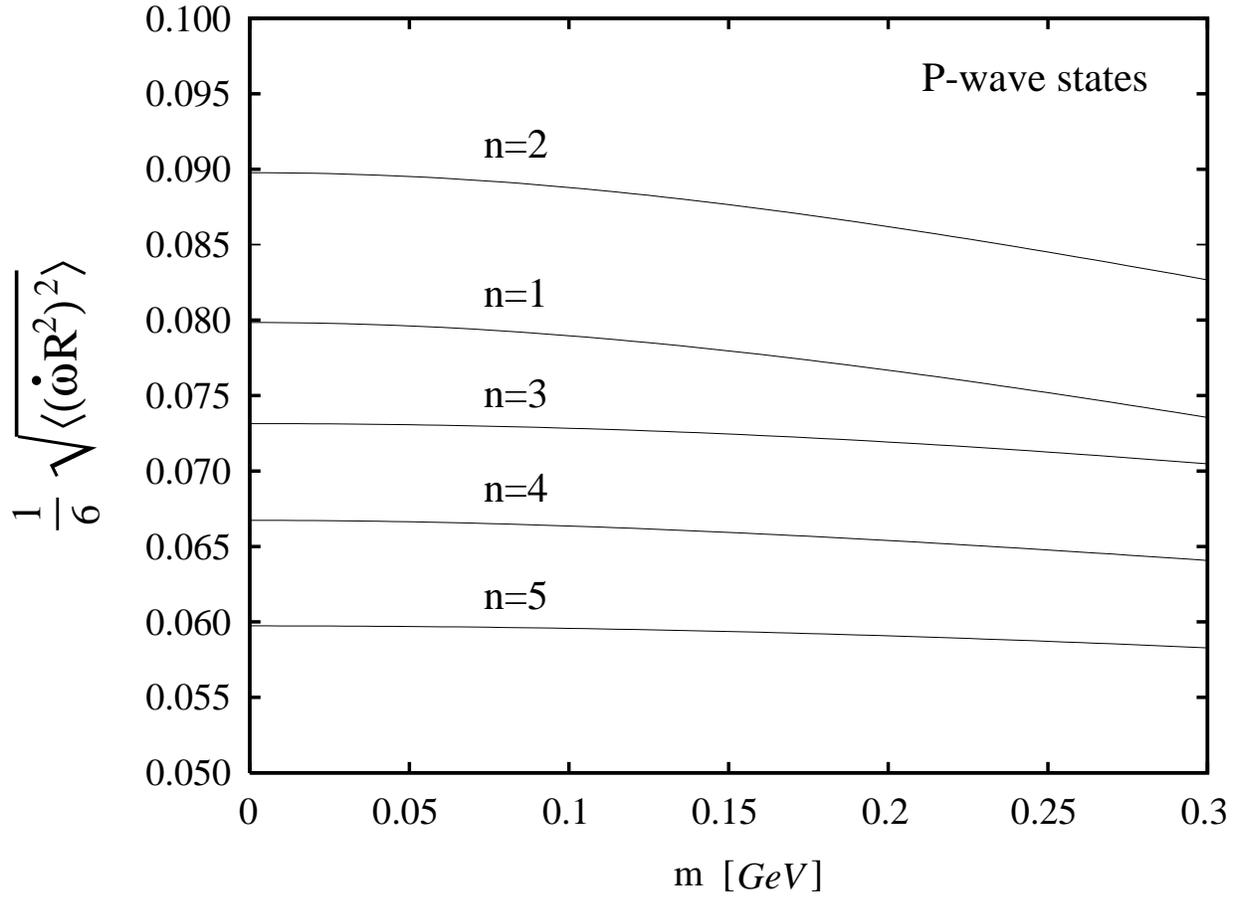}}
\caption{The dependence of ${1\over 6}\protect\sqrt{\langle (\dot{\omega}
R^2)^2 \rangle}$ on the light quark mass $m$.  These results were obtained
for the $P$-wave states in the heavy-light RFT model, with string tension
$a=0.2~GeV^2$. The $n=1$ line denotes the ground state, while the other
lines correspond to the first four radially excited
states.}\label{fig:mdep}
\end{figure}

\begin{figure}[p]
\vspace*{2.0in}
\epsfxsize=\hsize
\hbox{\hskip 0.0 in \epsfbox{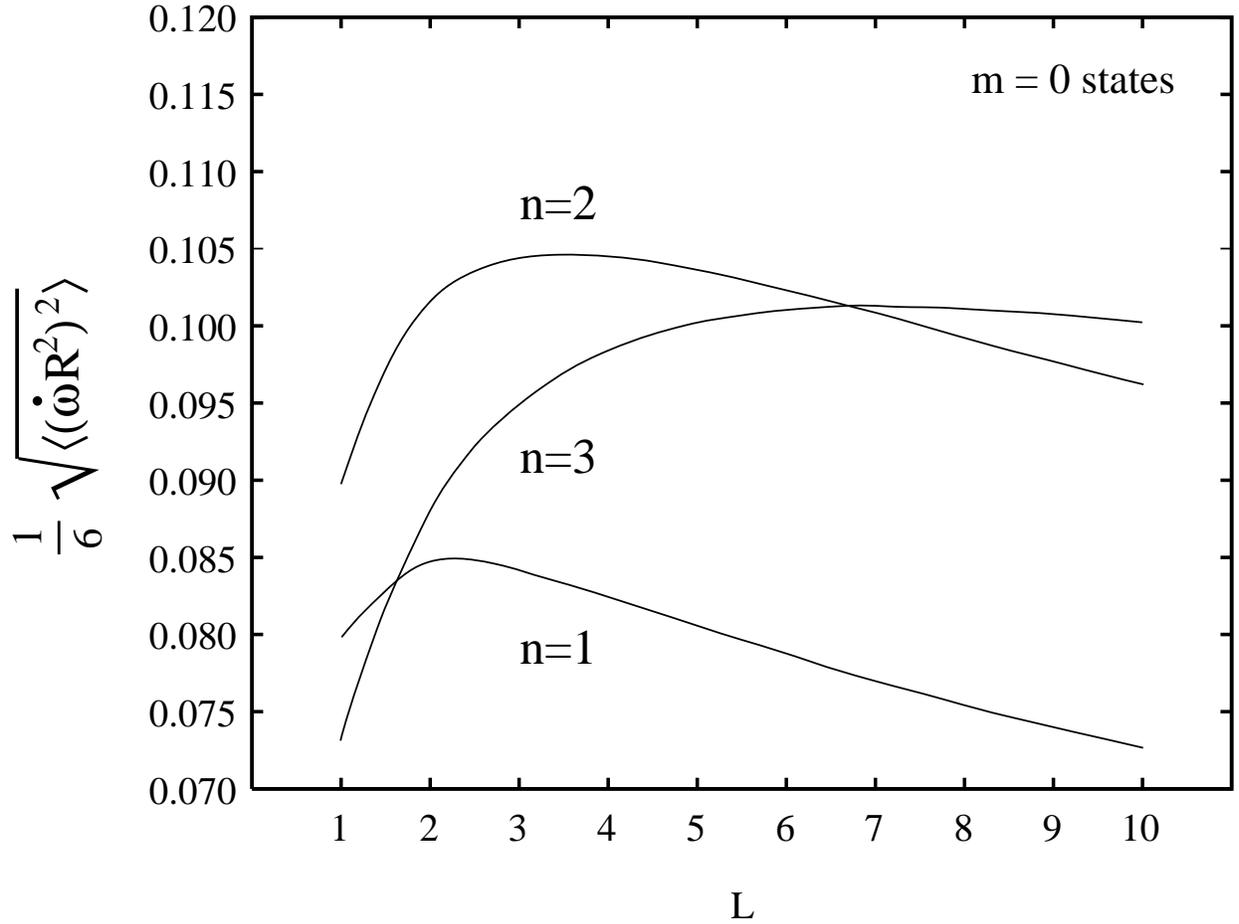}}
\caption{The dependence of ${1\over 6}\protect\sqrt{\langle (\dot{\omega}
R^2)^2 \rangle}$ on the angular momentum.  These results were obtained in
the heavy-light RFT model, with string tension $a=0.2\ GeV^2$ and light
quark mass $m=0$. The $n=1$ line corresponds to the ground state. Also
shown are the first two radially excited states.}\label{fig:ldep}
\end{figure}

Figures \ref{fig:mdep} and \ref{fig:ldep} show the dependence of the
quantity ${1\over6} \sqrt{\langle (\dot{\omega} R^2)^2 \rangle}$ on the
light quark mass and angular momentum, respectively. These results indicate
that the actual size of the displacement from the straight string in real
mesons is smaller than $10\%$, which illustrates the validity of the
straight string approximation.  Note that numerical estimates shown in
those figures were obtained using only the confining part of the
heavy-light RFT model with string tension $a=0.2\ GeV^2$.  Addition of the
short-range one gluon exchange interaction would further reduce our
results.

\section{String angular momentum, energy, and linear momentum}
\label{sec:constants}

We now compute the effects of the deformation away from
straightness on string dynamics.  To this end we compare the angular
momentum, energy, and linear momentum of the actual string with that of the
straight string.  The similarities and differences are quite interesting.
We proceed in each case by substituting the perturbed string form
(\ref{perturbX}) and the consequent perturbed metric into the desired
dynamical quantities.

\subsection{Angular momentum}

The expression (\ref{J}) for the angular momentum to first order in small
quantities becomes 
\begin{equation}
J^3 = J^3_{\rm straight} + {a R v_\perp\over \sqrt{1-v_\perp^2}}\, {\rm
Re}\,F(1) + {a \dot\phi R^3\over v_\perp^2}\left({1\over\sqrt{1-v_\perp^2}}
- {\arcsin v_\perp\over v_\perp} \right)\ , \label{Jfull}
\end{equation}
where we refer to Eq.~(\ref{Jstraight}) for the straight string result.  To
compare with the straight string angular momentum we must have strings of
the same length and with the same end velocity which requires that both
${\rm Re}\, F(1)$ and $\dot\phi$ vanish.  We then conclude that for small
deviations from straightness the angular momentum of the curved string is
the same as that of the straight string,
\begin{equation}
J^3 = J^3_{\rm straight}\  .
\end{equation}

\subsection{Energy}

Similarly, when we evaluate Eq.~(\ref{Pmu}) with $\mu = 0$ we obtain
\begin{equation}
E = E_{\rm straight} + {a\, {\rm Re}\, F(1) \over \sqrt{1-v_\perp^2}} - a\,
{\rm Re}\, F(0) + {a R^2 \dot\phi \over
v_\perp}\left({1\over\sqrt{1-v_\perp^2}} - {\arcsin v_\perp\over
v_\perp}\right)\  .
\end{equation}
Again requiring the accelerating string to have the same length and angular
velocity as the straight string, ${\rm Re}\, F(1)$ = ${\rm Re}\, F(0)$ =
$\dot\phi \equiv 0$, we have
\begin{equation}
E = E_{\rm straight}\ .
\end{equation}

\subsection{Linear momentum}

Finally we compute the linear momentum, 
\begin{equation}
P\equiv {1\over\sqrt2}\left(P_R + i P_\perp\right)\ ,
\end{equation}
of the curved string using
the spatial part of Eq.~(\ref{Pmu}).
The transverse momentum component is
\begin{equation}
P_\perp = P_{\perp\, \rm straight} + {a v_\perp\over\sqrt{1- v^2_\perp}} \,
{\rm Re}\, F(1) + {a\dot\phi R^2\over
v_\perp^2}\left({1\over\sqrt{1-v_\perp^2}} -1\right)\ .
\end{equation}
With the usual end conditions, we obtain
\begin{equation}
P_\perp = P_{\perp\, \rm straight}\ ,
\end{equation}
where $ P_{\perp\, \rm straight}$  is given in Eq.~(\ref{Pperp}).  The curved
string also has radial momentum
\begin{equation}
P_R = - a R^2\omega \int_0^1 d\sigma\, \sigma\gamma_\perp {df \over d\sigma}\ ,
\label{radP}
\end{equation}
which, after integration by parts, becomes
\begin{equation}
P_R = a R v_\perp\int_0^1 d\sigma\, {f(\sigma)\over (1-\sigma^2
v_\perp^2)^{3/2}}\ .
\end{equation}
It is worth noting that the expression for radial momentum (\ref{radP}) (and
also $P_\perp$) can be directly read off of Fig.~\ref{fig:one}.

Referring back to our explicit solution (\ref{shapefcn}) for $f(\sigma)$, we
find the analytic solution $P_R$ to be
\begin{equation}
P_R = -\,a R { \dot\omega R^2 \over v_\perp^4}\left[1 -
\sqrt{1-v_\perp^2} - {v_\perp\over 2}\arcsin v_\perp\right] \arcsin v_\perp\ .
\end{equation}
For small orbital velocities we have 
\begin{equation}
P_R \buildrel {v_\perp \ll 1} \over \longrightarrow -\, {a R \dot\omega R^2
v_\perp\over 24}\ .
\end{equation}

To understand the size of the radial string momentum we will compare it to
other ``relativistic corrections'' that arise in meson dynamics.  In a
meson with a quark mass $m$ large enough that the quark velocity is small,
the quark's angular momentum dominates that of the string.  In this case
$J\simeq m R^2\omega$ $\simeq$ constant and we have $R\dot\omega \simeq -
2\dot R \omega$.  The radial momentum is primarily due to the
non-relativistic quark with corrections from relativity and the string;
\begin{equation}
P_R^{\rm tot} = m\dot{R}\left[1 + {1\over 2} v_\perp^2 + {aR\over 12 m}
v_\perp^2 + \ldots\right]\ .
\end{equation}
The string radial momentum is smaller by a factor of $aR\over 6 m$ than the
first relativistic correction.

In this section we have observed that to leading order the angular
momentum, energy, and transverse momentum of the curved string are
unchanged by small deviations from a straight string.   The bending of the
string will induce a radial momentum but it is of higher order than the
leading relativistic corrections.  In this way, we agree with the work of
Brambilla et al. \cite{brambilla} who showed that in the Wilson loop
formalism relativistic corrections are correctly computed assuming a
straight path between the quarks.  

\section{Conclusions}\label{sec:conclusion}

We have considered here the shape of a QCD string with one end fixed and
the other moving with arbitrary velocity, but with a small angular
acceleration. We calculated the deviation from a straight string and found
it to reduce to a physically reasonable non-relativistic limit.  The
solutions we have obtained are adiabatic in the sense that they depend only
on the end condition, and would be static in the end rest frame.

To find the shape of the curved string, we perturbed an exact straight
solution to the Nambu-Goto-Polyakov string equations and solved exactly the
resulting equations of motion to leading order in the perturbation.  As
long as the angular acceleration is sufficiently small this should provide
an accurate picture of the string shape.  We numerically solved a straight
string model to estimate the angular acceleration that occurs in actual
mesons.  The result was that the string deviation from a straight line is
never very large, justifying the use of our perturbative approach, and
showing the validity of the straight string approximation.

In addition we have computed the angular momentum, energy, and linear
momentum of the curved string.  In each case but one, the perturbation
drops out and the perturbed string behaves identically to the straight one.
Only for the radial momentum does the deviation from straightness have an
effect.  In the semi-relativistic approximation this radial momentum is
smaller by a factor of ${aR\over6m}$ than the first relativistic
correction.  The straight string approximation is then justified for heavy
quark mesons as previously pointed out from a different point of view
\cite{brambilla}.  For relativistic mesons the string radial momentum may
have small, but perhaps interesting and calculable consequences.

\acknowledgments 
This work was supported in part by the US Department of Energy under
Contract No.~DE-FG02-95ER40896.  Fermilab is operated by URA under DOE
contract DE-AC02-76CH03000.

\appendix

\section{Conservation of momentum and angular momentum}\label{app:cons}

Here we consider the energy, momentum and angular momentum of a string with
a quark at each end.  The relevant actions are the Nambu-Goto-Polyakov
action, Eq.~(\ref{action}), and the quark action
\begin{equation}
S_{\rm quark} = {1\over 2}\sum_{i=1,2} \int d\tau\,  (e_{i}^{-1} \dot
x_{i}^\mu \dot x^{\phantom{\mu}}_{i\,\mu} - e^{\phantom{i}}_{i} m_{i}^2)\ , \label{quarkaction}
\end{equation}
in which $x_i^\mu$ is the position of the i$^{\rm th}$ quark of
mass $m_i$, and the auxiliary field $e_{i}$ is a one-dimensional metric
density analogous to the string's auxiliary field metric $h_{ab}$. The
boundary condition relating the two actions (\ref{action}) and
(\ref{quarkaction}) is that the quarks sit at the ends of the string,
\begin{equation}\label{BC}
x_{i}^\mu(\tau) = X^\mu(\tau,\sigma_i)\ ,\quad i = 1,2\ .
\end{equation}
Variation of the sum of the actions (\ref{action}) plus (\ref{quarkaction})
with respect to $X^\mu$ yields the equations of motion for the quarks
\begin{equation}\label{quarkEoMa}
\dot{p}_{i}^\mu = (-1)^i \,{\cal P}^{\mu\, \sigma}\Big|_{\sigma =
\sigma_i}\ ,
\end{equation}
where ${p}_{i}^\mu$ is the momentum of the quark at $\sigma_i = (-1)^i$,
related to the quark velocity by
\begin{equation}
p^\mu_i \equiv {\delta S_{\rm quark} \over \delta \dot{x}_{i\,\mu}}=
e^{-1}_i \dot{x}^\mu_i = m_i\, {\dot{x}^\mu_i \over \sqrt{-\dot{x}^\mu_i
\dot{x}_{{i\,\mu}}}}\ ,
\end{equation}
and 
\begin{equation}
{\cal P}^{\mu\, a} \equiv {\delta S\over \delta X_{\mu,a}} = -\,
a \sqrt{-h}\,h^{ab}\, X^\mu\mathstrut_{,b}
\end{equation}
is the momentum current on the string worldsheet, which is related to the
canonical string momentum $\Pi^\mu$ by $\Pi^\mu = {\cal
P}^{\mu\, \tau}$. 

It is easy to show that the total momentum of the string--quark system,
\begin{equation}
P^\mu = \int_{\sigma_1}^{\sigma_2}d\sigma\, {\cal P}^{\mu\, \tau} 
+ \sum_{i=1,2} p^\mu_{i}\ ,
\end{equation}
is conserved under $\tau$ evolution,
\begin{eqnarray}
\partial_\tau{P}^\mu &=& \int_{\sigma_1}^{\sigma_2}d\sigma\, \dot{{\cal
P}}^{\mu\, \tau} + \sum_{i=1,2} \dot{p}^\mu_{i} \nonumber \\ &=&
-\int_{\sigma_1}^{\sigma_2}d\sigma\, {{\cal P}}^{\mu\, \sigma\, \prime} +
\sum_{i=1,2} \dot{p}^\mu_{i}\nonumber \\ &=& \dot{p}^\mu_{1} + {\cal
P}^{\mu\, \sigma}\Big|_{\sigma = \sigma_1} + \dot{p}^\mu_{2} - {\cal
P}^{\mu\, \sigma}\Big|_{\sigma = \sigma_2}\ =\ 0 \ . \label{momcons}
\end{eqnarray}
The equations of motion ${\cal P}^{\mu\, a}\mathstrut_{,a} = 0$ are
used to go from the first to the second line, and the final result vanishes
by the quark equations of motion (\ref{quarkEoMa}).

The conservation of the angular momentum,
\begin{equation}
J^{\mu\nu} = \int_{\sigma_1}^{\sigma_2}d\sigma\, X^{[\mu}{\cal P}^{\,\nu]\,
\tau} + \sum_{i=1,2} x_{i}^{[\mu}p^{\nu]}_{i}\ ,
\end{equation}
is only slightly more complicated to demonstrate. The $\tau$ derivative of
$J^{\mu\nu}$ is
\begin{eqnarray}
\partial_\tau{J}^{\mu\nu} &=& \int_{\sigma_1}^{\sigma_2}d\sigma\,
\left(\dot{X}^{[\mu}{\cal P}^{\,\nu]\, \tau} + X^{[\mu} \dot{{\cal
P}}^{\,\nu]\, \tau}\right) + \sum_{i=1,2}
\left(\dot{x}_{i}^{[\mu}p^{\nu]}_{i} +
x_{i}^{[\mu}\dot{p}^{\nu]}_{i}\right) \nonumber \\ &=&
\int_{\sigma_1}^{\sigma_2}d\sigma\, \left(\dot{X}^{[\mu}{\cal P}^{\,\nu]\,
\tau} - X^{[\mu}{{\cal P}}^{\,\nu]\, \sigma\,\prime}\right) + \sum_{i=1,2}
x_{i}^{[\mu}\dot{p}^{\nu]}_{i}\nonumber \\ &=&
\int_{\sigma_1}^{\sigma_2}d\sigma\, \left(-{X}^{[\mu\,\prime} {\cal
P}^{\,\nu]\, \sigma} - X^{[\mu}{{\cal P}}^{\,\nu]\, \sigma\,\prime}\right)
+ \sum_{i=1,2} x_{i}^{[\mu} \dot{p}^{\nu]}_{i}\nonumber \\ &=& -{X}^{[\mu}{\cal
P}^{\,\nu]\, \sigma} \Big|_{\sigma=\sigma_2} + X^{[\mu} {\cal P}^{\,\nu]\,
\sigma} \Big|_{\sigma=\sigma_1} + \sum_{i=1,2}
x_{i}^{[\mu}\dot{p}^{\nu]}_{i}\ .
\end{eqnarray}
In going from the second to the third line, we have used $\dot{X}^{[\mu}
{\cal P}^{\nu]\, \tau} = - a \sqrt{-h}\, h^{\tau a}\, \dot{X}^{[\mu}
{X}^{\nu]}\mathstrut_{,a} = - a \sqrt{-h}\, h^{\tau \sigma}\,
\dot{X}^{[\mu} {X}^{\nu]\,\prime} = -{X}^{[\mu\,\prime} {\cal P}^{\,\nu]\,
\sigma}$.  Using the boundary conditions (\ref{BC}) that place the quark at
the end of the string, we find
\begin{eqnarray}
\partial_\tau{J}^{\mu\nu} &=& \left(x_{1}^{[\mu}\dot{p}^{\nu]}_{1} 
+ x_{1}^{[\mu}{\cal P}^{\nu]\, \sigma}\Big|_{\sigma=\sigma_1}\right) +
\left(x_{2}^{[\mu}\dot{p}^{\nu]}_{2} - x_{2}^{[\mu}{\cal P}^{\nu]\,
\sigma}\Big|_{\sigma=\sigma_2}\right) = 0\ .\label{angmomcons} 
\end{eqnarray}

Finally, the quark mass-shell condition, 
\begin{equation}
p_{i}^\mu p_{i\,\mu}  + m_{i}^2 = 0\ ,
\end{equation}
which follows identically from the relation of the quark momenta to quark
velocities,
\begin{equation}
p^\mu_i =  m_i {\dot{x}^\mu_i \over \sqrt{-\dot{x}^\mu_i
\dot{x}_{i\,\mu}}}\ ,
\end{equation}
must be preserved by the equations of motion (\ref{quarkEoM}).  We find that
\begin{eqnarray}
\partial_\tau\left(p^\mu p_\mu + m^2 \right) &=& 2p^\mu\dot{p}_\mu \nonumber \\
&\propto& \ p^\mu h^{\sigma a} X_{\mu,a} \ \propto \ \dot{X}^\mu h^{\sigma
a} X_{\mu,a}\nonumber \\ 
&=& h^{\sigma a} X_{\mu,a} X^\mu\mathstrut_{,\tau} = h^{\sigma a} 
h_{a\tau} = \delta^\sigma_\tau = 0\ . \label{shellcons}
\end{eqnarray}

The three conditions (\ref{momcons}), (\ref{angmomcons}), and
(\ref{shellcons}) together imply that the conservation of energy momentum,
angular momentum, and the quark mass-shell relation are equivalent to the
quark equations of motion, Eqs.~(\ref{quarkEoM}) and (\ref{quarkEoMa}).

\section{Exact solution of the relativistic string shape
equation}\label{app:diffeq}

We consider here the details of the solution of (\ref{shapediffeq})
\begin{equation}
(1-\xi^2) {d^2 f\over d\xi^2} + \xi {df\over d\xi} - f = \beta\, \xi 
\end{equation}
to obtain the central result (\ref{shapefcn}) of the paper.  The solution
method follows standard procedures.  First we define a new function, $g$,
so that 
\begin{equation}
f(\xi) = \xi g(\xi)\ ,
\end{equation}
giving
\begin{equation}
\xi(1-\xi^2) g^{\prime\prime} + (2-\xi^2) g^\prime = \beta\,\xi\ .
\end{equation}
Now define
\begin{equation}
g^\prime \equiv  H(\xi) I(\xi) \label{eqn:a4}
\end{equation}
to obtain
\begin{equation}
\xi(1-\xi^2) I(\xi) H^\prime + H\left[\xi(1-\xi^2) I^\prime + (2-\xi^2)
I\right] = \beta\,\xi\ . \label{eqn:a5}
\end{equation}
The integrating factor $I(\xi)$ is chosen to make the coefficient of $H$
vanish and by quadrature,
\begin{equation}
I(\xi) = {\sqrt{1-\xi^2}\over \xi^2}\ .
\end{equation}
Substituting into (\ref{eqn:a5}), we find 
\begin{equation}
H(\xi) = \beta\left({\xi\over\sqrt{1-\xi^2}} - \arcsin\xi\right) + C_2\ ,
\end{equation}
and, by (\ref{eqn:a4}), 
\begin{equation}
{d g\over d\xi} = \beta\left({1\over \xi} -
{\xi\over\sqrt{1-\xi^2}}\arcsin\xi\right) + C_2\left(\sqrt{1-\xi^2} +
\xi\,\arcsin\xi\right)\ .
\end{equation}
Finally by quadrature we have $g(\xi)$ and $f=\xi g$ is
\begin{equation}
f(\xi) = \beta \left(\sqrt{1-\xi^2} + {\xi\over 2} \arcsin\xi 
\right)\arcsin\xi + C_1 \xi + C_2 \left(\sqrt{1-\xi^2} +
\xi\,\arcsin\xi\right)\ .
\end{equation}

\newpage

\newpage

\newpage

\end{document}